\documentclass[prl,twocolumn,superscriptaddress,showpacs]{revtex4}
\usepackage{graphicx}
\usepackage{dcolumn}
\usepackage{bm}

\begin{document}
\title{Dynamic Alignment in Driven Magnetohydrodynamic Turbulence}
\author{Joanne Mason} 
\author{Fausto Cattaneo}
\affiliation{Department of Astronomy and Astrophysics, University of Chicago, 
5640 South Ellis Ave, Chicago, IL 60637}
\author{Stanislav Boldyrev}
\affiliation{Department of Physics, University of Wisconsin at Madison, 1150 University Ave, 
Madison, WI 53706}
\date{\today}

\begin{abstract}
Motivated by recent analytic predictions, we report numerical evidence showing that in 
driven incompressible magnetohydrodynamic turbulence the
magnetic- and velocity-field fluctuations locally tend to align the directions of their  
polarizations. This dynamic alignment is stronger 
at smaller scales with the angular mismatch between the 
polarizations decreasing with the scale $\lambda$ approximately 
as $\theta_{\lambda}\propto \lambda^{1/4}$. 
This can naturally lead to a weakening of the nonlinear interactions and provide an 
explanation for the energy 
spectrum $E(k)\propto k^{-3/2}$  that is observed in numerical 
experiments of strongly magnetized turbulence. 
\end{abstract}

\maketitle

\section{Introduction.}
Incompressible hydrodynamic turbulence is 
described by the Kolmogorov theory of energy cascade, in which turbulent fluctuations (or eddies) transfer energy from 
the largest scales, where it is injected, to smaller 
eddies until the dissipative scale is reached and energy is extracted from the system (e.g., \citep{frisch,biskamp}).  
Such an energy cascade is local, i.e. its rate at each scale 
depends solely upon the characteristics of eddies at that scale. 
It follows that an eddy of size $\lambda$, with typical velocity fluctuations of strength $\delta v_{\lambda}$, will 
lose energy to (or get fragmented into) smaller-scale eddies over a time of duration $\tau_{\lambda}\propto 
\lambda/\delta v_{\lambda}$ (the only quantity having the dimensions of time that can be constructed in this system).
In a steady state,  the energy 
flux from the injection scale to the dissipative scale is the same at 
all scales, leading to $\delta v_{\lambda}^2/\tau_{\lambda}=const$ and $\delta v_{\lambda}\propto \lambda^{1/3}$. The 
corresponding Fourier spectrum of velocity fluctuations then has the Kolmogorov scaling, $E_k \propto k^{-5/3}$. 

Iroshnikov \citep{iroshnikov} and Kraichnan \citep{kraichnan} realized 
that magnetohydrodynamic (MHD) turbulence is qualitatively different from 
non-magnetized turbulence. The governing equations of incompressible MHD are (see, e.g., \cite{biskamp})
\begin{eqnarray}
\label{eq:mhd}
\partial_t {\bf v}+({\bf v}\cdot \nabla){\bf v}=-\nabla p+ \left(\nabla \times {\bf B}\right) \times {\bf B}+\nu 
\Delta {\bf v}+{\bf f},\nonumber \\
\partial_t {\bf B} = \nabla \times ({\bf v} \times {\bf B})+\eta \Delta {\bf B},\\
\nabla \cdot {\bf v}=0, \quad \nabla \cdot {\bf B}=0, \nonumber
\end{eqnarray}
where ${\bf v }({\bf x}, t)$ is the velocity field, ${\bf B }({\bf x}, t)$ the magnetic field, $p$ the pressure and 
${\bf f}({\bf x},t)$ is the external force.
Unlike the large-scale velocity which can be removed from the hydrodynamic system by means of a Galilean 
transformation (thus allowing the hydrodynamic system to be treated locally), here the magnetic field of large-scale 
eddies cannot be eliminated by transforming into a moving reference frame. Thus the small-scale eddies always 
experience the action 
of the large-scale magnetic field. The MHD turbulent cascade is therefore mediated 
by such a guiding field. 

The Iroshnikov-Kraichnan theory of MHD turbulence is formulated upon the assumption that the turbulence 
is three-dimensionally isotropic. 
In this theory, the 
turbulence becomes progressively weaker as the cascade proceeds to smaller scales 
and the turbulent spectrum has the form $E_{k}\propto k^{-3/2}$. A detailed discussion  
of the Iroshnikov-Kraichnan theory can be found, for example, in \cite{biskamp}. 

The isotropy of turbulence in the presence of a strong, large-scale magnetic field was questioned 
in early analytic and numerical considerations (see, e.g., \cite{Shebalin,Strauss,Montgomery,Kadomtsev}), 
and it was demonstrated that the energy cascade is directed mostly perpendicularly to the guiding field (e.g., 
\cite{Goldreich94}).  In 1995,  
Goldreich and Sridhar \cite{goldreich} developed a theory by proposing that 
the turbulent eddies become progressively more stretched along the guiding field 
as the cascade proceeds toward small scales. As a result, 
the time of nonlinear interaction can be estimated as in the Kolmogorov theory, 
however, the energy cascade is anisotropic with respect to the guiding field. The {\em field-perpendicular} 
energy spectrum then has then the form $E(k_{\perp}) \propto k_{\perp}^{-5/3} $. 

In recent years, increasingly high-resolution numerical simulations have indeed confirmed the 
anisotropy of MHD turbulence. The turbulent fluctuations are elongated 
along the guiding field (see, e.g., \citep{milano,cho,maron,biskamp-muller}), and the anisotropy is scale-dependent. 
However, the numerical simulations also find the field-perpendicular energy spectrum $E(k_{\perp}) \propto 
k_{\perp}^{-3/2}$ (\cite{maron,biskamp-muller,muller}; see also \cite{haugen}). Thus the findings are neither 
described by the isotropic Iroshnikov-Kraichnan theory, nor do they agree with the 
Goldreich-Sridhar scaling for the energy spectrum. 

A possible resolution to this controversy has been recently proposed in \citep{boldyrev,boldyrev2}. 
To discuss it, let us first note that at each wave number ${\bf k}$, the Fourier components of fluctuating 
fields can be expanded in shear-Alfv\'en and pseudo-Alfv\'en modes (we will provide more detail 
in the next section). Since the turbulent cascade 
proceeds mainly in the field perpendicular direction, it is dominated by shear-Alfv\'en modes, 
while the pseudo-Alfv\'en modes are passively advected by the turbulent cascade 
(see \citep{goldreich} for a detailed discussion). 
It was suggested in \citep{boldyrev,boldyrev2}  
that the polarizations of the shear-Alfv\'en magnetic-field and velocity-field fluctuations 
become spontaneously aligned in a turbulent cascade. 
This alignment becomes progressively stronger for smaller scales.  
It was proposed that at a given scale $\lambda$ 
the typical fluctuations $\delta {\bf v}_{\lambda}$ and $\pm \delta {\bf b}_{\lambda}$ are 
aligned within the angle 
\begin{eqnarray}
\theta_{\lambda}\propto \lambda^{1/4}.
\label{alignment}
\end{eqnarray} 
As in the Goldreich-Sridhar theory, the 
eddies are stretched along the guiding field. However, as a result of polarization alignment, they 
also become anisotropic in the field-perpendicular plane, and this anisotropy increases as the 
scale decreases. This leads to a scale-dependent depletion of the nonlinear interaction in~(\ref{eq:mhd}) 
and to the spectrum 
of field-perpendicular fluctuations $E(k_{\perp})\propto k_{\perp}^{-3/2}$, in good agreement with 
the numerical results described above.

As proposed in \cite{boldyrev,boldyrev2}, 
the reason for such a dynamic alignment may be the existence of {\em two} conserved quantities 
in magnetohydrodynamics, whose cascades are directed toward small scales in a 
turbulent state. The MHD system (\ref{eq:mhd}) in the ideal limit $\nu=\eta=0$ conserves the integral 
of energy  
\begin{eqnarray}
E=\frac{1}{2}\int (b^2+v^2)d^3 x,
\label{energy}
\end{eqnarray}
and the integral of cross-helicity,
\begin{eqnarray}
H^C=\int ({\bf v}\cdot {\bf b})d^3 x,
\label{crosshelicity}
\end{eqnarray}
provided that the fluctuations~${\bf v}({\bf x})$ 
and~${\bf b}(\bf{x})$ have periodic 
boundary conditions or vanish at infinity. It was 
proposed  in \citep{boldyrev,boldyrev2} that the sole requirement of constant energy 
flux does not allow one to find 
the spectrum of the fluctuations uniquely. The structure of turbulent fluctuations in 
the inertial region should accommodate constant fluxes of both conserved quantities, which 
leads to scale-dependent anisotropy of turbulent eddies in the field perpendicular plane, 
and to the alignment~(\ref{alignment}).   

Interestingly, the effect of dynamic alignment has been extensively investigated in the case of  
{\em decaying} MHD turbulence (e.g. \citep{dobrowolny,grappin,pouquet,pouquet_sm,politano}), where it essentially 
means that decaying magnetic and velocity 
fields approach asymptotically in time the so-called Alfv\'enic state ${\bf v}(x)=\pm {\bf b}(x)$. However, decaying 
turbulence is qualitatively 
different from its forced counterpart, and the effects that we discuss in the 
present work have not been addressed in the earlier investigations.   

We also mention that previous explanations of the numerically observed 
shallower-than-Kolmogorov spectrum have essentially invoked intermittency 
effects, e.g., \citep{maron,biskamp-muller,beresnyak-lazarian}. Although intermittency may   
significantly affect the scaling of 
higher-order correlation functions, it usually provides only small 
corrections to the energy spectrum. Our explanation of the -3/2 spectrum 
does not require intermittency. 

In the present paper we investigate the phenomenon 
of dynamic alignment via direct numerical simulations. We analyze 
driven incompressible MHD turbulence with a strong guiding magnetic field. 
We measure the degree to which the velocity 
and magnetic field fluctuations align as a function of scale and we also investigate the dependence on 
the strength of the imposed field. Numerical verification of the scale-dependent dynamic 
alignment (\ref{alignment}) is the main goal of our work.

\section{Dynamic alignment in MHD turbulence.}
\label{dynamicalignment}

We solve the MHD equations (\ref{eq:mhd}) using standard pseudospectral methods. An external magnetic field is applied 
in $z$ direction with 
strength $B_0$ measured in units of velocity. The periodic domain has a resolution of $256^3$ mesh points and is 
elongated in the $z$ direction, with aspect ratio 1:1:$B_0$. The external force,  ${\bf f} ({\bf x},t)$, is chosen so 
as to drive the turbulence at large scales and it satisfies the following requirements: it has no component along $z$, 
it is solenoidal in the $x-y$ plane, all the Fourier coefficients outside the range $1 \leq k \leq 2$ are zero, the 
Fourier coefficients inside that range are Gaussian random numbers with amplitude chosen so that the 
resulting {\it rms} velocity fluctuations are of order unity, and the individual random values are refreshed 
independently {\it on average} every turnover time of the large scale eddies. 
The Reynolds number is defined as $Re=U_{rms}L/\nu$, 
where $L$ $(\sim 1)$ is the field-perpendicular box size, $\nu$ is fluid viscosity, 
and $U_{rms}$ $(\sim 1)$ is the rms 
value of velocity fluctuations. 
We restrict ourselves to the case in which magnetic resistivity and fluid viscosity are the same, $\nu=\eta$. The 
system is evolved until a stationary state is reached (confirmed by observing the time evolution of the total energy 
of fluctuations) and the data are then sampled in intervals of order of the eddy turn-over time. All results presented 
correspond to averages over these samples (approx. 10 samples).

\begin{figure}[tbp]
\vskip-5mm
\centerline{\includegraphics[width=3.7in]{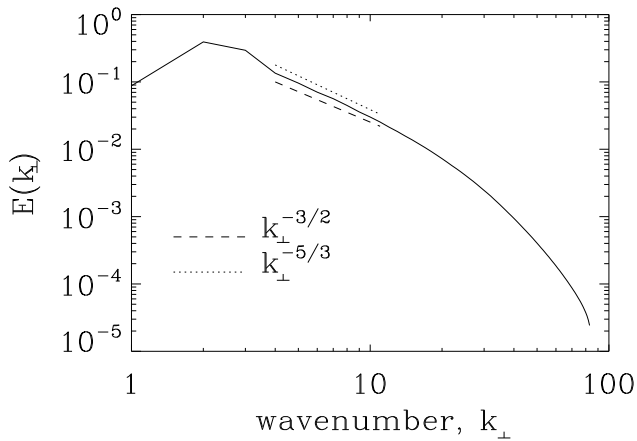}}
\vskip-5mm
\caption{\label{spectrum800} Two-dimensional spectrum of MHD turbulence in the plane perpendicular to 
the guiding field ${\bf B}_0$. The numerical 
resolution is $256^3$, the Reynolds number is $Re=800$, and the strength of the uniform guiding 
field is $B_0=5$. The turbulence is driven by a solenoidal random force at wave numbers $k=1,2$, as is explained in 
the text.}
\end{figure}

First we measure the two-dimensional energy spectrum, defined as $E(k_{\perp})=\langle |{\bf v}(k_{\perp})|^2\rangle 
k_{\perp} +\langle|{\bf b}(k_{\perp})|^2\rangle k_{\perp}$, 
where ${\bf v}(k_{\perp})$ and ${\bf b}(k_{\perp})$ are two-dimensional Fourier transformations 
of the velocity and magnetic fields in a plane perpendicular to ${\bf B}_0$ and $k_\perp=\left(k_x^2+k_y^2 
\right)^{1/2}$. The average is taken over all such planes in the data cube, and then over all data cubes. The 
resulting spectrum of fluctuations is presented in Fig.~\ref{spectrum800}. It is impossible to infer the 
exponent of the power-law distribution with good accuracy here and in particular it is hard to distinguish between the 
spectral indices $-5/3$ and $-3/2$. As was demonstrated in \cite{biskamp-muller,muller}, a much higher resolution 
allows greater Reynolds numbers to be explored and yields a longer inertial range. Since recent high resolution 
calculations do yield $E(k_{\perp}) \sim k_{\perp}^{-3/2}$ (see  \citep{maron,biskamp-muller,muller,haugen}) we choose 
not to pursue this issue further here. Instead we concentrate our study on the effect of dynamic alignment that, as we 
shall demonstrate presently, can be observed well even with limited resolution. 

To investigate the dynamic alignment of the shear-Alfv\'en fluctuations in the field-perpendicular plane, we use 
specially constructed structure functions. Let us denote $\delta {\bf v}_{r}={\bf v}({\bf x}+{\bf r})-{\bf v}({\bf 
x})$ and $\delta {\bf b}_{r}={\bf b}({\bf x}+{\bf r})-{\bf b}({\bf x})$, where ${\bf r}$ is a 
point-separation vector in a plane perpendicular to the large-scale field, ${\bf B}_0$ (see \footnote{ 
The characteristic scale $\lambda$ discussed in the introduction refers to those $r=|{\bf r}| $ 
that are representative 
of the inertial range.}). In incompressible MHD there are  two types of linear modes: shear-Alfv\'en waves 
and pseudo-Alfv\'en waves. Consider a wave whose wave vector ${\bf k}$ is almost 
perpendicular to the guiding field. Then the shear-Alfv\'en wave will have polarization 
of ${\bf v}$ and ${\bf b}$ fluctuations in the direction perpendicular to 
both ${\bf B}_0$ and ${\bf k}$. The polarization of the pseudo-Alfv\'en wave will lie in 
the plane of ${\bf B}_0$ and ${\bf k}$, perpendicularly to ${\bf k}$. Since the wave  
vector is almost perpendicular to ${\bf B}_0$, the polarization of the pseudo-Alfv\'en
 wave will be closely aligned with ${\bf B}_0$. 

In the nonlinear case, the Fourier component of fluctuations at each 
wave vector ${\bf k}$ can be expanded into two components: those with polarizations in the 
${\bf B}_0\times {\bf k}$ direction, and those in the $\left({\bf B}_0\times {\bf k}\right)\times {\bf k}$ 
direction. Although, in contrast with the linear case, such modes are generally not 
solutions of the MHD equations, we will refer to them as the shear-Alfv\'en mode and  
the pseudo-Alfv\'en mode, respectively. As we mentioned earlier (see also \citep{goldreich}),  
the shear-Alfv\'en modes play the dominant role in the anisotropic cascade, 
while the pseudo-Alfv\'en modes are passively advected by them. The scaling of 
the pseudo-Alfv\'en spectrum then follows the scaling of the shear-Alfv\'en 
spectrum. However, since the pseudo-Alfv\'en modes do not dominate  the 
dynamics, they should be excluded when the angular alignment is calculated. 
In other words, in order to restrict ourselves to shear-Alfv\'en fluctuations we need to 
exclude the components of $\delta {\bf v}_{r}$ and $\delta {\bf b}_{r}$ 
parallel to the large scale magnetic field. 

It is important to note however, that since the turbulence is strong an eddy 
of size $r$ lives only one eddy turn-over, during which time it can be transported along the large scale field only by
a distance comparable to its own size. 
Therefore, during its life time, such an eddy feels only the local 
direction of the guiding field which can be tilted with respect to the direction of the global  
field ${\bf B}_0$ by the action of larger and longer lived eddies; this was recognized by Cho \& Vishniac~\citep{cho}.   
Therefore, when calculating the angular alignment we need to remove the components of $\delta {\bf v}_{r}$ and $\delta 
{\bf b}_{r}$ that are in the direction of the 
{\em local} magnetic field~${\bf B}({\bf x})$ (when ${\bf B}_0$ is strong, ${\bf B}({\bf x})\approx {\bf B}_0$).  
Thus we calculate $\delta {\tilde {\bf v}}_{r}=\delta {\bf v}_{r}-(\delta 
{\bf v}_{r}\cdot {\bf n}){\bf n}$ and $\delta {\tilde {\bf b}}_{r}=\delta {\bf b}_{r}-(\delta {\bf b}_{r}\cdot {\bf 
n}){\bf n}$, where ${\bf n}={\bf B}({\bf x})/\vert {\bf B}({\bf x})\vert$.  

We are now ready to introduce the following two structure 
functions. The first  is defined by
\begin{eqnarray}
S_{cross}(r)=\langle \vert \delta {\tilde {\bf v}}_{r}\times \delta {\tilde{\bf b}}_{r} \vert \rangle,
\end{eqnarray}
where $\times$ denotes a vector cross-product, and the average is performed over all different positions of the point 
${\bf x}$. The second 
structure function is defined by
\begin{eqnarray}
S_2(r)=\langle \vert \delta {\tilde{\bf v}}_{r} \vert \vert  \delta {\tilde{\bf b}}_{r} \vert \rangle,
\end{eqnarray}
with the same averaging procedure. By definition of the cross product, the two structure functions are related by the 
angle between the polarizations 
of $\delta {\tilde{\bf v}}_{r}$ and $\delta {\tilde{\bf b}}_{r}$. This is precisely the phenomenon we would like to 
investigate.  

\begin{figure} [tbp]
\vskip-5mm
\centerline{\includegraphics[width=3.9in]{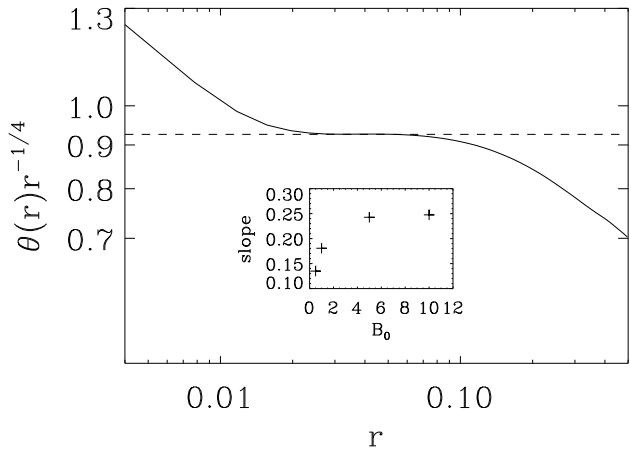}}
\vskip-5mm
\caption{The alignment angle, defined in Eq.~(\ref{angle}), compensated by the theoretical scaling $r^{1/4}$ ($B_0=5, 
Re=800$). The insert demonstrates how the slope $d\theta/dr$ in the inertial range changes with the strength of the 
guiding field.}
\label{alignment800}
\end{figure}

When the alignment angle $\theta_{r}$, say, is small, we have 
\begin{eqnarray}
\theta_{r} \approx \sin \left(\theta_r \right) \equiv S_{cross}(r)/S_2(r).
\label{angle}
\end{eqnarray}
Presented in Fig.~\ref{alignment800} is the result of such a numerical calculation. There the angle is compensated by 
the phenomenological scaling $r^{1/4}$ of \citep{boldyrev,boldyrev2}. The data shows a clear good agreement with the 
theory from approximately $1/50$ to $1/10$ of the field-perpendicular box size. The effects of varying the guiding 
field strength are shown in the insert of Fig.~\ref{alignment800}. We note that  provided that the field is 
sufficiently strong the alignment effect is robust and does not change significantly with field strength. For weaker 
fields on the other hand, the alignment is gradually 
lost.

We should note that
recently, Beresnyak and Lazarian \cite{beresnyak-lazarian} attempted to 
measure the geometric alignment predicted in \cite{boldyrev,boldyrev2} 
by a procedure analogous to averaging the angle between the {\em normalized} fluctuations 
$\delta {\hat {\bf v}}_{r}=\delta {\bf v}_{r}/|\delta {\bf v}_{r}|$ and
$\delta {\hat {\bf b}}_{r}=\delta {\bf b}_{r}/|\delta {\bf b}_{r}|$.    
They did not observe the alignment~(\ref{alignment}) and concluded that the weakening of 
interaction was, rather, an intermittency effect.   
Their averaging procedure, however, does not respect the fact 
that, by definition,  the $\delta {\bf v}_\lambda$ 
and $\delta {\bf b}_\lambda$ to which (\ref{alignment}) refers are typical fluctuations 
at the scale $\lambda$, that is, fluctuations whose strengths
are close to their rms values (cf. \cite{frisch,biskamp}). The alignment angle should 
be measured between such {\em dynamically relevant} fluctuations, as, 
for instance, in our approach~(\ref{angle}). 

The numerical verification 
of the scale dependent dynamic alignment of the magnetic and velocity polarizations 
in driven MHD turbulence, presented in Fig.~\ref{alignment800}, is the main result of our work.

\section{Discussion}
Two consequences of the observed polarization alignment have particular importance.  
These concern the energy spectrum and the viscous scale (or inner scale) of MHD turbulence. 

{\em Energy spectrum.}---The scale dependent alignment can naturally imply that  
the energy transfer time is $\tau_{\lambda}\sim \lambda/(\delta v_{\lambda}\theta_{\lambda})$ 
(see \citep{boldyrev,boldyrev2}). Since we 
obtained $\theta_{\lambda}\propto \lambda^{1/4}$,  
the requirement of constant energy flux, $\delta v^2_{\lambda}/\tau_{\lambda}={\rm const}$, 
then leads to $\delta v_{\lambda}\propto \lambda^{1/4}$, where $\lambda$ is the field-perpendicular scale of 
fluctuations. This translates to the field-perpendicular  
energy spectrum of MHD turbulence, $E(k_{\perp}) \propto k_{\perp}^{-3/2}$, 
announced in the introduction.  

{\em Viscous scale.}---At the viscous scale $\lambda_{\nu}$, the time of nonlinear interaction $\tau_{\lambda}$ 
is of the order of  the diffusive time $\tau_{\nu}\sim \lambda^2/\nu$. A simple 
calculation then leads to $\lambda_{\nu}\sim L/{\rm Re}^{2/3}$. This result is 
qualitatively different from that for nonmagnetized turbulence, $\lambda_{\nu}(B=0)\sim 
L/{\rm Re}^{3/4}$ (see e.g., \cite{frisch}). This implies that for the same Reynolds 
numbers, the viscous scale in MHD turbulence is larger than that in nonmagnetized 
turbulence. This difference is especially relevant for astrophysical plasmas, where 
Reynolds numbers are quite large, $\rm Re\sim 10^5-10^{10}$. 

Apart from its fundamental value, the existence of dynamic alignment in driven MHD turbulence 
has consequences for our understanding of such astrophysical phenomena as solar wind structure, 
interstellar scintillation, cosmic-ray transport in galaxies, and heat conduction in galaxy 
clusters. A discussion of these matters will be presented elsewhere.

\acknowledgments
We acknowledge the hospitality of the Aspen Center for 
Physics where part of this work was carried out. This work was supported by the NSF Center for Magnetic
Self-Organization in Laboratory and Astrophysical Plasmas
at the Universities of Chicago and Wisconsin-Madison.

\end {document}